\documentclass[aip,rsi,reprint,groupedaddress,english]{revtex4-1}
\usepackage[pdfborder={0 0 0}]{hyperref}
\usepackage{color}
\usepackage[usenames,dvipsnames]{xcolor}
\usepackage{soul}

\usepackage{amsmath}
\usepackage{amsfonts}
\usepackage{amssymb}
\usepackage{graphicx}
\usepackage{array}
\usepackage{epstopdf}

\begin{document}

\title{A three-dimensional steerable optical tweezer system for ultracold atoms}

\author{C. S. Chisholm} \altaffiliation[Current address: ]{ICFO--Institut de Ciencies Fotoniques, The Barcelona
Institute of Science and Technology, 08860 Castelldefels
(Barcelona), Spain.}
\author{R. Thomas}
\author{A. B. Deb}
\author{N. Kj{\ae}rgaard}
\email[Corresponding author: ]{niels.kjaergaard@otago.ac.nz}
\affiliation{Department of Physics, QSO--Centre for Quantum Science, and Dodd-Walls Centre for Photonic and Quantum Technologies, University of Otago, New Zealand.}

\date{\today}

\begin{abstract}

We present a three-dimensional steerable optical tweezer system based on two pairs of acousto-optic deflectors. Radio frequency signals used to steer the optical tweezers are generated by direct digital synthesis and multiple time averaged cross beam dipole traps can be produced through rapid frequency toggling. We produce arrays of ultracold atomic clouds in both horizontal and vertical planes and use this as to demonstrate the three-dimensional nature of this optical tweezer system.

\end{abstract}

\maketitle

\section{Introduction}
\label{sec:introduction}

Optical trapping has been a valuable tool in the field of ultracold atoms since the first demonstration by Chu~\textit{et al.}~\cite{Chu1986}. In particular, optical potentials are widely used due to being largely state insensitive~\cite{Grimm2000} which leaves quantum spin as a degree of freedom~\cite{McGuirk2002} and allows the use of external magnetic fields for tuning atom-atom interactions via Feshbach resonances~\cite{Horvath2017,Sawyer2017,Cabrera2017}. The introduction of the cross beam dipole trap, which provides tight confinement in three dimensions~\cite{Adams1995}, has enabled a number of advances such as all optical production of Bose-Einstein condensates (BECs)~\cite{Barrett2001}; notably, condensation of cesium is not possible in a magnetic trap due to unfavorable scattering lengths~\cite{Weber2002}. Since then, the generation of arbitrary and time dependent optical potentials has been demonstrated~\cite{Henderson2009}. Such potentials are desirable for use in the fields of atomtronics~\cite{Seaman2007,Pritchard2012}, atom interferometry~\cite{Sherlock2011}, and atom based quantum simulation~\cite{Hayward2012,Fialko2017,Sturm2017} and magnetic field gradiometry~\cite{Wood2015}. To this end, a variety of techniques for engineering dynamically configurable optical potential have been employed including the use of liquid-crystal spatial light modulators~\cite{Grier2006,Boyer2006,Kim2016,Barredo2016,Haase2017}, digital-micromirror devices~\cite{Muldoon2012,Preiss2015,Gauthier2016}, rapidly toggled \cite{Henderson2009,Roberts2014,Bell2016,Bell2018a} and multi-toned acousto-optic deflectors (AODs)~\cite{Zawadzki2010,Zimmermann2011,Rakonjac2012,Trypogeorgos2013,Endres2016}, and cavity modes of optical resonators\cite{Naik2017}.

In this article, we present a configuration of four AODs, arranged in two orthogonal pairs, with the fundamental diffraction orders of each pair forming a cross beam dipole trap~\cite{Adams1995}. This layout, paired with rapid toggling of the AOD driving frequencies~\cite{Henderson2009,Roberts2014,Bell2016}, allows the simultaneous steering of multiple dipole traps in three-dimensional space. We show that such a configuration can be used to create a variety of non-trivial potentials in one-, two-, and three-dimensions. Specifically, we demonstrate $2\times2$ arrays of atomic clouds on both horizontal and vertical planes. Potential future avenues for our three-dimensional tweezer system include vortex formation and superfluid flow by merging of independent BECs~\cite{Scherer2007,Aidelsburger2017}, study of vortex lattice decay to quantum turbulence through collisions of multiple BECs in nonlinear geometries~\cite{Mawson2015}, and quantum optics with Rydberg states of atoms~\cite{Reinhard2007,Beguin2013,Firstenberg2016,Busche2017}.

\section{Design}
\label{sec:design}

To construct our optical tweezer system, we used two pairs of orthogonal AODs (AAoptoelectronic, DTSXY-400). Each pair is aligned for $+1$ order Bragg diffraction on both AODs~\cite{Magdich1989} and operated with nominal center acoustic frequencies of $75~\text{MHz}$. Light sourced from a $1064~\text{nm}$, $50~\text{W}$ single frequency fiber laser (IPG Photonics, YLR-50-1064-LP-SF) is delivered to the AODs using photonic crystal fibers (PCFs)\footnote{NKT Photonics, LMA-PM-15}. The Radio frequency signals driving the AODs are generated by direct digital synthesis (DDS) using a commercially available device (Wieserlabs, FlexDDS-NG Rack).

\begin{figure}
\includegraphics{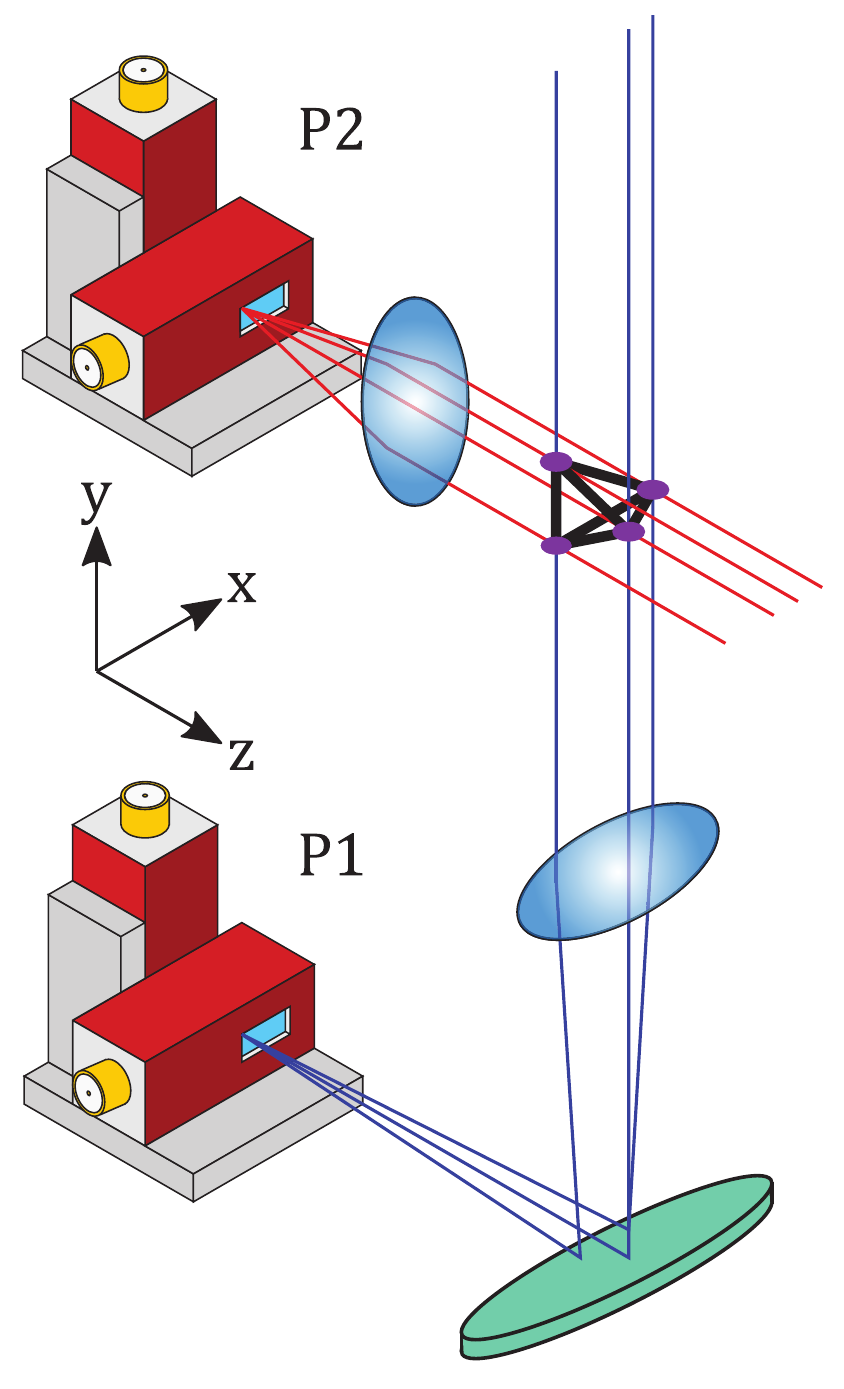}
\caption{\label{fig1} Simplified arrangement of two pairs (P1 and P1) of orthogonal AODs used to implement a three-dimensional optical tweezer system. Beams from AOD pair P1 propagate vertically (blue lines) and beams from AOD P2 propagate horizontally (red lines). Purple ellipses represent an example of a 3D configuration of atomic clouds established by four cross beam dipole traps at the vertices of a tetrahedron. The thick black lines joining the tetrahedrally arranged atomic clouds are to assist with three-dimensional.}
\end{figure}

Figure~\ref{fig1} shows a simplified arrangement of our optical tweezer system. The pairs of AODs are arranged such that, at the nominal center driving frequency, the doubly diffracted beam from one pair (P1) propagates along the $y$-axis and can be steered in the $xz$-plane by dynamically altering the driving frequency in either one or both of AODs in the pair. Similarly, the doubly diffracted beam from the second pair (P2) propagates along the $z$-axis and can be steered in the $xy$-plane. The nominal spot size radius at beam waist for the doubly diffracted beam from boths pairs of AODs is $40~\mu\text{m}$. The vertical axis is aligned to the direction of acceleration due to gravity and is labeled as the $y$-axis.  Multiple time averaged beams can be produced by repeatedly changing the AOD driving frequencies~\cite{Henderson2009,Roberts2014,Bell2016}. The toggled beams are made to be parallel to each other using lenses placed approximately one focal length in front of the AODs. In Fig.~\ref{fig1} an example of four cross beam dipole traps in a tetrahedral arrangement is shown.

Up to $7.6\pm0.5~\text{W}$ of light can be delivered to the AODs via each of the PCFs. The diffraction efficiency at the center frequencies of both pairs of AODs was measured to be $\sim73~\%$ with a full width at half maximum bandwidth greater than $25~\text{MHz}$ for all four AODs. The frequency change to displacement calibrations for P1 and P2 are $211\pm2~\mu\text{m}\cdot\text{MHz}^{-1}$ and $-397\pm3~\mu\text{m}\cdot\text{MHz}^{-1}$ respectively. The difference in scaling is attributed to the difference in focal lengths of the lenses used to correct propagation directions. The accessible volume for the optical tweezers exceeds $6~\text{mm}\times6~\text{mm}\times6~\text{mm}$, which is comparable to the volume defined by the Rayleigh range of the tweezer beams.

\begin{figure*}
\includegraphics{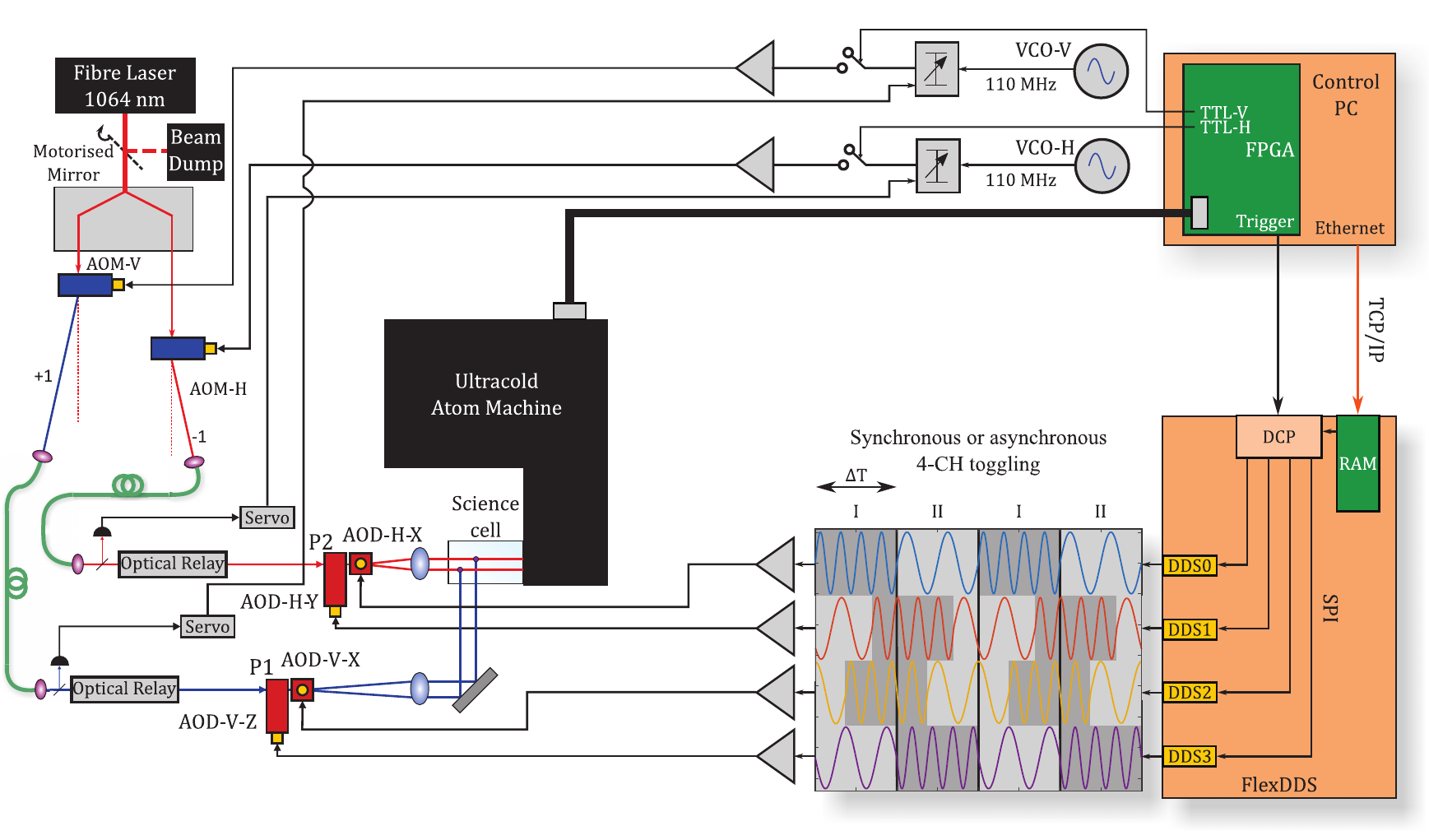}
\caption{\label{fig2} Schematic representation of the three-dimensional optical tweezers system. The optical output from a 1064~nm fibre laser providing light for the tweezers can be turned off by deflecting it into a beam dump with a motorized mirror. In its on-state light is split into two beam paths (V and H) using an arrangement of polarizing beam splitter cubes, wave plates and lenses summarized in a functional box. Acousto-optic modulators (AOM-V and AOM-H, respectively) diffract the V and H-beam into  PCFs linking the light source to the experiment. The PCF outputs are relayed to the AOD pairs steering the optical tweezers. An arrangement (combined into functional blocks in the schematic) of wave plates, lenses, and mirrors are used to optimize the polarization, size and collimation of the V- and H-beams for maximum diffraction efficiency. The AODs are driven by the amplified output of a FlexDDS unit (see text). Gray shading behind the waveforms delivered by the FlexDDS represents segments of constant frequency and black vertical bars represent arrival of trigger pulses from an FPGA (triggering of individual channels can be synchronous or asynchronous). The dwell time of the frequency hopping, $\Delta T$, which is equal to the inverse of the FlexDDS trigger rate is indicated on the figure.}
\end{figure*}

Figure~\ref{fig2} shows a schematic of the optical power delivery system and rf control. Fixed frequency acousto-optic modulators (AOMs) driven by amplified voltage controlled oscillators (VCOs) are used to rapidly switch and regulate the optical power incident on the AODs. The AOMs are aligned to the $+1$ order Bragg diffraction and the $-1$ order Bragg diffraction for P1 and P2, respectively, which prevents static interference between vertical and horizontal beams by ensuring frequency differences in excess of $160~\text{MHz}$, which are much faster than any dynamics we consider. In addition, the polarizations for the vertical and horizontal beams are chosen to be orthogonal. The optical power delivered to each AOD pair of is regulated independently by measuring a fraction of the output power\footnote{We use SUPERTHIN\textsuperscript{TM} (thickness $\sim 200$~$\mu m$) R/T 1/99 plate beam splitters (Spectral Optics part \# SSTPBS-1064-1-45P/S) to tap off light for servoing} transmitted by the PCFs and dynamically adjusting the amplitudes of the AOM rf drives. If we use only the AOMs to switch on and off trapping light, they will spend only a small fraction of time ($\lesssim 1\%$) in the on-state. During the on-time, the rf drive heats up the AOM crystals and causes the pointing of the diffracted beams to drift, compromising the coupling of light into the PCFs. In addition, exposing the AOM crystals to high optical power for long periods of time can lead to thermal lensing\cite{Bogan2015} and a consequent degradation of the beam mode and fiber coupling efficiency. To mitigate these issues, we use a motorized flipper mirror\footnote{Thorlabs MFF101 Motorized Filter Flip Mount (flip time $\lesssim 1$~s)} to divert the laser beam into a beam dump and keep the AOMs in the on-state during all preparatory stages of the experimental cycle. In this fashion, the AOMs are fed rf power more than 98\% of the time so that the crystal can reach an equilibrium temperature while only being exposed to optical power $\lesssim 2\%$ of the time\cite{Chisholm2018}.

 The FlexDDS unit generating the AOD rf drives can store more than $10^8$ instructions in RAM distributed between $2$ to $12$ channels, depending on the configuration. A single instruction can set the phase, amplitude and frequency of a single channel. Typically one time step for optical tweezers consisting of two AODs requires four instructions, including wait-for-trigger commands. Each channel can be addressed individually and has a sample rate of $1~\text{GS}\cdot\text{s}^{-1}$ with a frequency range from $0.3~\text{Hz}$ to $400~\text{MHz}$. The FlexDDS enables control of the acoustic power inside the AODs via amplitude tuning words, which allows dynamic balancing of power between toggled beams. DDS instructions are loaded into RAM on the FlexDDS via TCP/IP and subsequently written to DDS chips via serial peripheral interface (SPI) by a digital command processor (DCP). The timing for the execution of FlexDDS instructions is controlled using external triggers from a field programmable gate array (FPGA). Instructions can be triggered at rates in excess of $250~\text{kHz}$ which are limited by the SPI write time of the FlexDDS and the access time of the AODs~\cite{Bell2016}. Additionally, individual time delays of up to $134~\text{ms}$ with $8~\text{ns}$ resolution can be specified for each output channel (the master clock is locked to a GPS-disciplined $10~\text{MHz}$ quartz oscillator reference). These time delays are used to account for finite propagation time of the acoustic waves when the frequencies are toggled, which result in ``ghost'' traps when toggling two or more acousto-optic devices in series.

\begin{figure}
\includegraphics{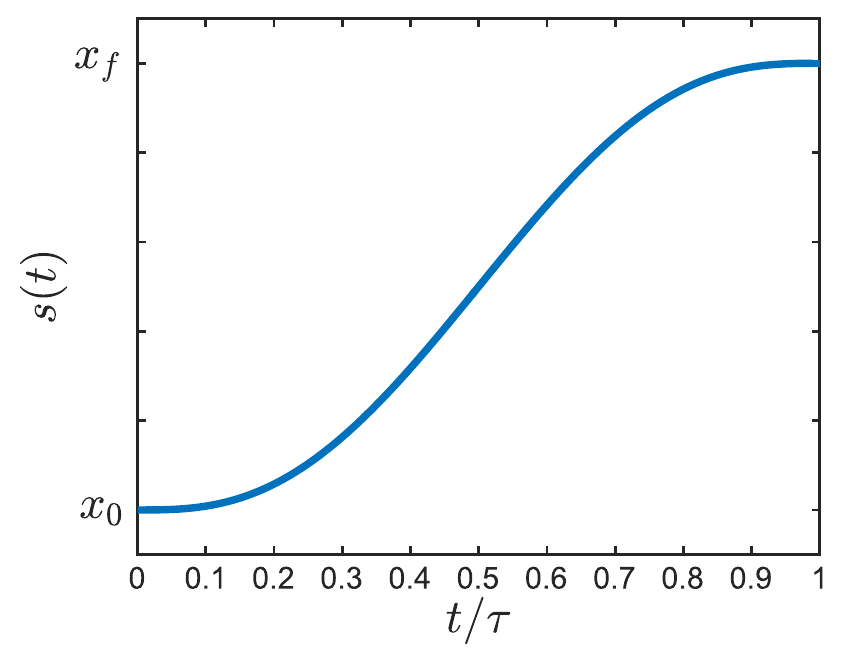}
\caption{\label{fig3} An example of a minimum-jerk cost trajectory, as defined by Eq.~\ref{eq1}, in one dimension.}
\end{figure}

In our experiment, ultracold $^{87}$Rb atoms in the $5^2\text{S}_{1/2}|F=2,m_F=2\rangle$ ground state are loaded into a cross beam dipole trap from a Ioffe-Pritchard (IP) trap. The initial dipole trap typically consists of one static horizontal beam (propagating along the $z$-axis of the IP trap) and two rapidly toggled vertical beams separated by $60~\mu\text{m}$~\cite{Deb2014}. Atomic clouds can be evaporatively cooled to quantum degeneracy by controlled decrease of the optical power in the horizontal beam(s)~\cite{Adams1995,Barrett2001}. To produce multiple clouds, we pull tweezer beams apart on minimum-jerk cost trajectories before evaporation. A minimum-jerk cost trajectory minimizes the integral of the square of the third derivative (the jerk) of the trajectory~\cite{Flash1985,McKellarMSc}
\begin{equation*}
\int_0^{\tau}\left[\frac{\mathrm{d}^3}{\mathrm{d}t^3}\mathbf{s}(t)\right]^2\mathrm{d}t,
\end{equation*}
where we have assumed that the total transit time is $\tau$. By applying boundary conditions of $\mathbf{s}(0)=\mathbf{x}_0$, $\mathbf{s}(\tau)=\mathbf{x}_f$ and zero initial and final velocity and acceleration, we see that a minimum-jerk cost trajectory is defined by
\begin{equation}
\mathbf{s}(t) = \mathbf{x}_0 + \mathbf{D}\left[10\left(t/\tau\right)^3 - 15\left(t/\tau\right)^4 + 6\left(t/\tau\right)^5\right],
\label{eq1}
\end{equation}
where $\mathbf{D} = (\mathbf{x}_f - \mathbf{x}_0)$. An example of a minimum-jerk cost trajectory in one-dimension is shown in Fig.~\ref{fig3}.

\begin{figure*}
\includegraphics[width=\textwidth]{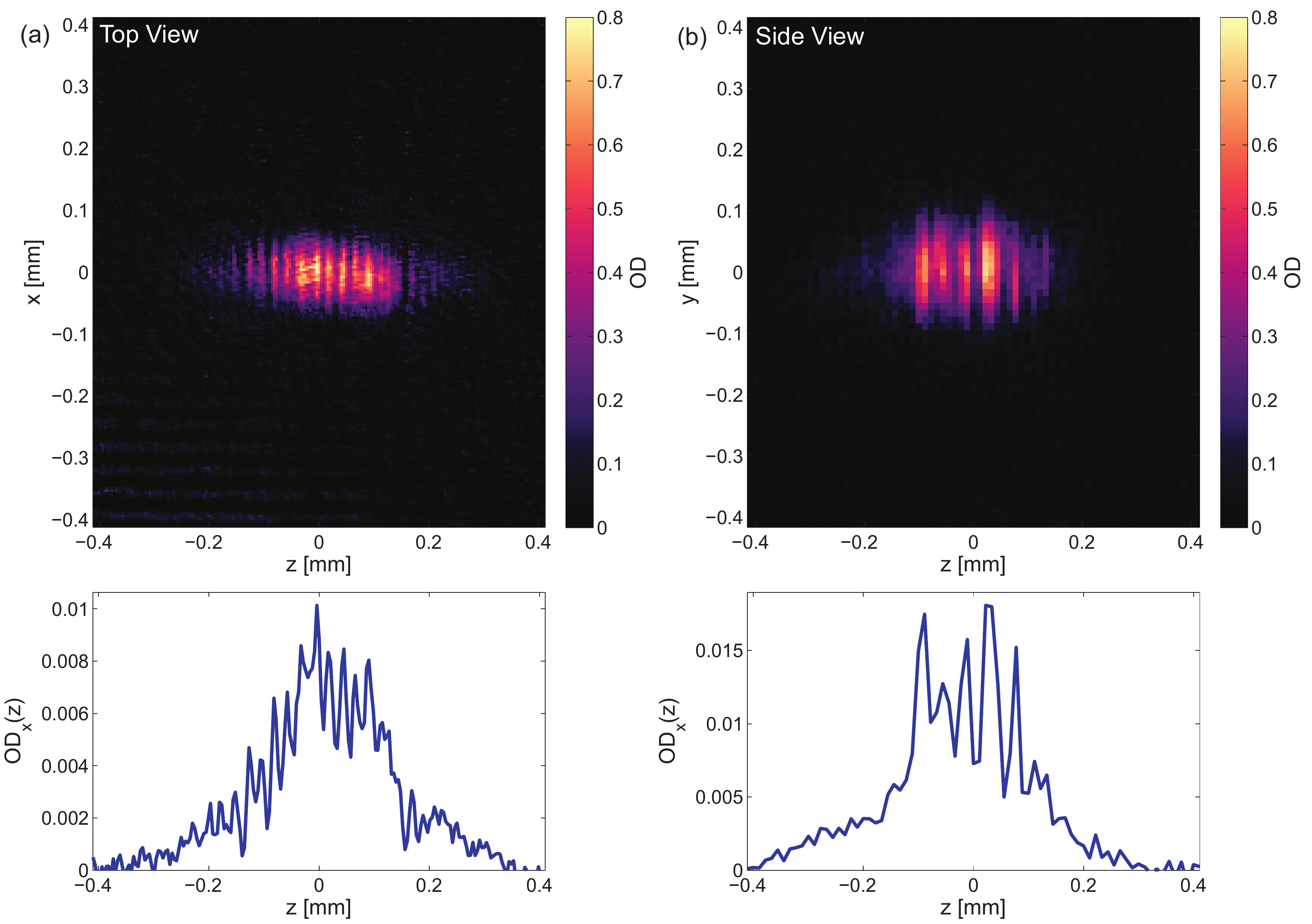}
\caption{\label{fig4} (a) Matter wave interference fringes between two BECs of $^{87}$Rb with an initial separation of $80~\mu\text{m}$, observed with the vertical imaging path at $30~\text{ms}$ time-of-flight. The experimental sequence is described in the main text. The color bars represent optical depth. (b) Matter wave interference fringes, produced using the same method as in (a), captured at $50~\text{ms}$ time-of-flight using the horizontal imaging path. The lower panels show the optical depth integrated with respect to $x$, $\mathrm{OD}_x(z)$ over a three-pixel wide region.}
\end{figure*}

\section{Imaging}
\label{sec:imaging}
\begin{figure*}
\includegraphics{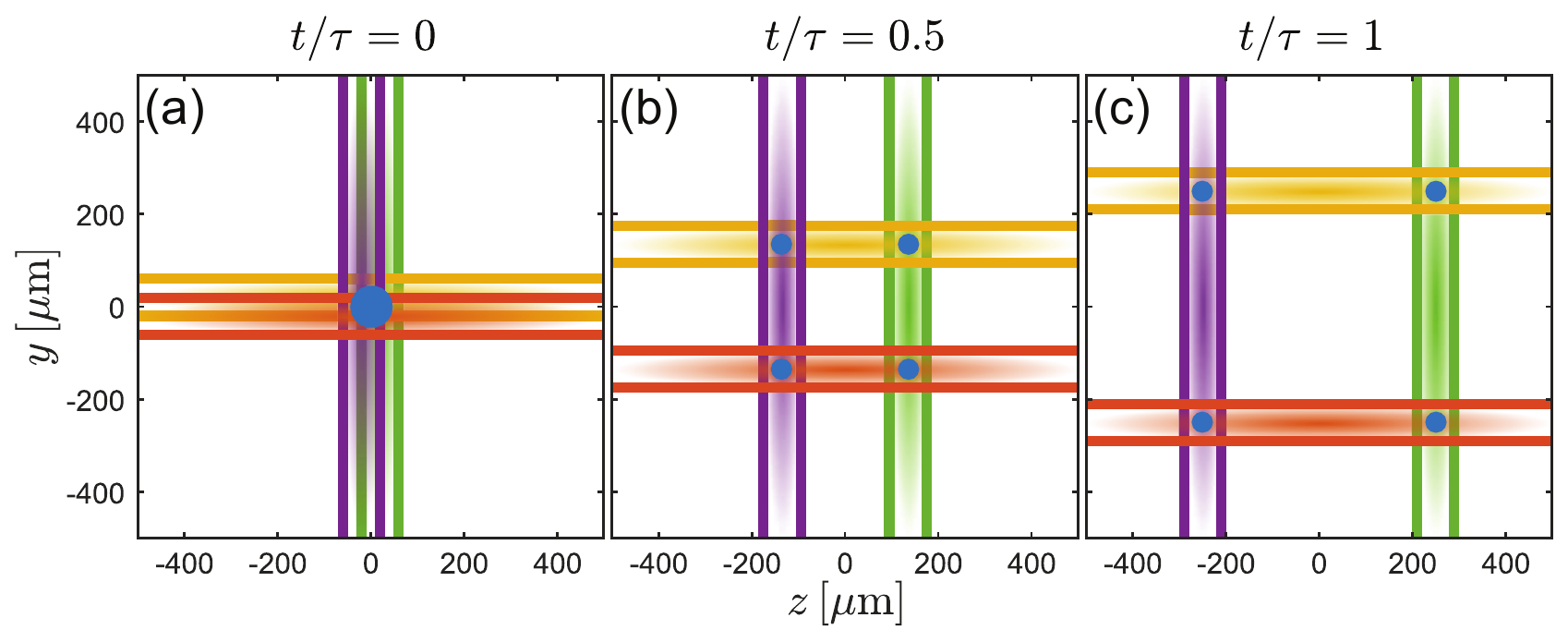}
\caption{\label{fig5} An illustration of an experimental sequence which produces four atomic clouds in a square arrangement with nearest neighbor separation of $500~\mu\text{m}$. The laser beams are false colored for clarity.}
\end{figure*}
Our experimental platform was built with two orthogonal absorption imaging paths for addressing atoms in the science cell. Both paths image $^{87}$Rb via the $\text{D}_2$ line ($780~\text{nm}$). One path has the probe beam propagating along the $x$-axis (horizontal) towards a charge coupled device (CCD). The second imaging path has the probe beam propagating along the $y$-axis (vertical) towards a CCD. The pixel size limited resolutions of the horizontal and vertical imaging paths are $11.10\pm0.01~\mu\text{m}$ and $4.10\pm0.01~\mu\text{m}$, respectively.

The vertical imaging system also employs a current carrying coil located above the science cell which is used to levitate the atoms by applying a $15~\text{G}\cdot\text{cm}^{-1}$ field gradient~\cite{Li2015}. After adiabatic rapid passage to the strong field seeking state $|F=2,m_F=-2\rangle$, the levitation field is applied during ballistic expansion of the atoms in order to keep the atoms from falling away from the focus of the objective lens. The objective lens for the vertical imaging path is a diffraction limited singlet (Thorlabs, AL2550H) with a numerical aperture of $\text{NA} = 0.2$ which, according to the Rayleigh criterion, gives a diffraction limited resolution of $4.8~\mu\text{m}$.

To investigate the resolution of our imaging system along the two axes (horizontal and vertical) we inspected the matter wave interference between two BECs\cite{Kohstall2011,Zhai2018}. BECs were prepared in two cross beam dipole traps, separated along the $z$-axis by $80~\mu\text{m}$, consisting of a single horizontal beam and two rapidly toggled vertical beams (one for each trap). The optical power shared by the two vertical beams was ramped linearly to zero over a period of $100~\text{ms}$ after after which the two condensates were allowed to expand in the horizontal beam for $50~\text{ms}$ before being released into $30~\text{ms}$ time-of-flight under the influence of magnetic levitation. Figure~\ref{fig4}a shows an absorption image recorded along the vertical axis that presents the emergence of a clear interference pattern with a fringe contrast approaching 50\%. A comparable fringe contrast was not found for the horizontal imaging path using the same preparation sequence; rather a $50~\text{ms}$ time-of-flight was required to achieve a similar contrast (see Fig.~\ref{fig4}b).  The lower panels of Fig.~\ref{fig4} show the optical depth integrated with respect to $x$, $\mathrm{OD}_x(z)$. A simple peak finding algorithm gives fringe spacings of $23~\mu\text{m}$ for the vertical imaging path and $41~\mu\text{m}$ for the horizontal imaging path. We note that there are $\sim5$ fringes visible for the horizontal path and more than $10$ fringes visible for the vertical path. We attribute the apparent reduction in number of fringes of increased spacing to aliasing resulting form the horizontal imaging resolution being comparable to the fringe spacing. Altogether our results clearly demonstrate a higher spatial resolution for the vertical path compared with the horizontal path.

\section{Results}
\label{sec:results}

In this section, we present some preliminary demonstrations of our ability to manipulate arrays of atomic clouds in three-dimensions. As previously mentioned, to arrange multiple ultracold atomic clouds into arrays, we start with a single cross beam dipole trap of thermal atoms loaded from an IP trap. The thermal atoms are split by dynamically altering the trapping potential such that beams are split as necessary. All tweezer beams followed minimum-jerk cost trajectories. When powers were balanced correctly (particularly important when splitting along the axis of gravity), splitting a single beam into $N$ beams would result in the trapped atomic cloud being split into $N$ atomic clouds with approximately equal atom number.

Figure~\ref{fig5} illustrates how we produce four atomic clouds in a square array with nearest neighbor separation of $500~\mu\text{m}$. The cross beam dipole trap used for loading atoms from the IP trap is formed from four overlapping beams, two horizontal and two vertical, which are pulled apart on minimum-jerk cost trajectories. Frames (a) - (c) show sequentially increasing time with equal increments. In a typical experimental sequence the total time for splitting is on the order of $100~\text{ms}$.

\begin{figure}
\includegraphics{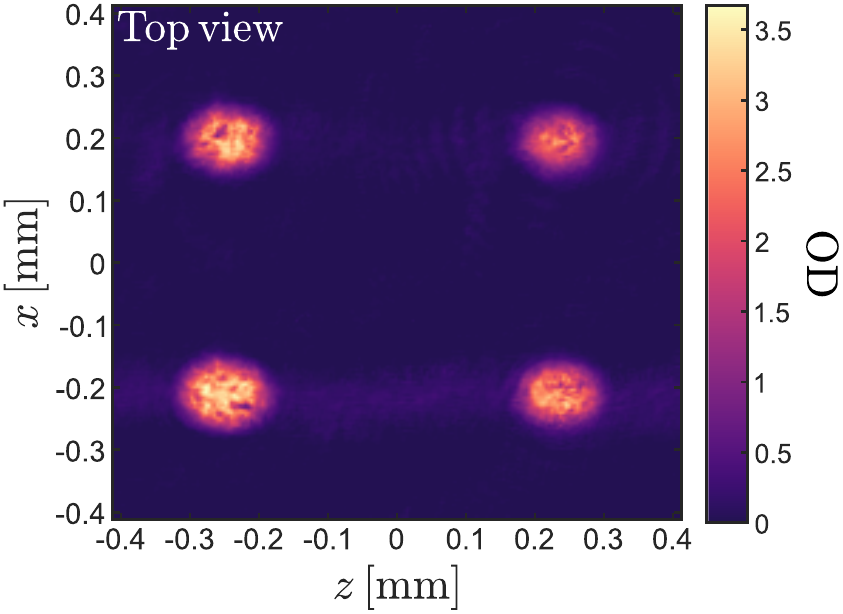}
\caption{\label{fig6} An array of thermal clouds in the $xz$-plane with all six beams used to form the four cross beam dipole traps generated by rapidly toggling AODs. The nearest neighbor separations were $\Delta\!\,x = 400~\mu\text{m}$ and $\Delta\!\,z = 500~\mu\text{m}$ and the time-of-flight was 2 ms. Due to this short time of flights the absorption image becomes saturated at the center or each cloud.}
\end{figure}

Figure~\ref{fig6} shows an absorption image captured using the vertical imaging system corresponding to four thermal clouds arranged in a horizontal plane. The arrangement was rectangular; the distance between neighboring clouds with respect to the $x$-axis was $400~\mu\text{m}$ and the separation with respect to the $z$-axis was $500~\mu\text{m}$. .

\begin{figure}
\includegraphics{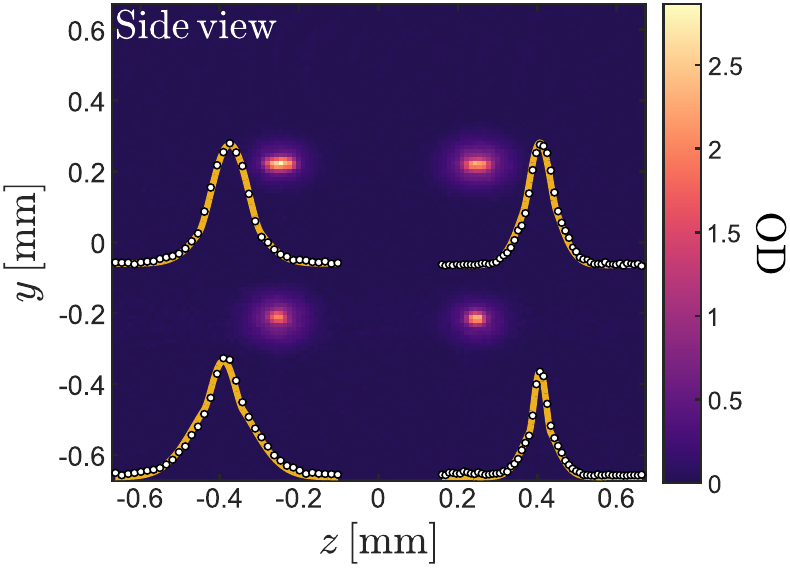}
\caption{\label{fig7} Four BECs arranged in a square in the $yz$-plane with nearest neighbor separation of $500~\mu\text{m}$. This image was taken at 12 ms time-of-flight. The summed projections of both the data and fits, shown below and to the sides of the atomic clouds (not to scale), demonstrate the characteristic bimodal density profile of BECs.}
\end{figure}

Creating a square arrangement in a vertical plane required a modification of the experimental sequence due to the influence of gravity. Most importantly, the power balance was adjusted to be heavily in favor of the upper beam since gravity would keep all of the atoms in the lower beam if the powers were equal. Second, the clouds were split in two steps. In the first step, clouds were split along the $y$-axis by $150~\mu\text{m}$ in $500~\text{ms}$. In the second step the clouds were simultaneously split by a further $350~\mu\text{m}$ in the $y$-direction and $500~\mu\text{m}$ in the $z$-direction in $100~\text{ms}$. Figure~\ref{fig7} shows an arrangement of clouds produced using the above experimental sequence. In addition, the clouds were evaporatively cooled over a period of $2~\text{s}$ to produce one BEC at each of the four sites.

At the end of evaporation, each of the four clouds shown in Fig.~\ref{fig7} contained more than $6\times10^4$ atoms and the condensate fractions were between $0.1$ and $0.4$. The condensate fractions were extracted from two-dimensional bimodal (projected parabolic, corresponding to a Thomas-Fermi condensate and Gaussian corresponding to the thermal component~\cite{Ketterle1999}) fits to the absorption image. The absorption image was acquired at $12~\text{ms}$ time-of-flight.

\section{Discussion}
\label{sec:discussion}
In the above, we have presented realizations of two-dimensional manipulation and arrangements of ultracold atoms on two orthogonal planes (horizontal and vertical) which \textit{ipso facto} demonstrates the three-dimensional capabilities of our steerable optical tweezer platform. Arguably, a more compelling demonstration would have included a truly 3D configuration of atomic clouds such as, e.g. the tetrahedron of Fig.~\ref{fig1}. A catastrophic breakdown of our fiber laser used to produce  light for the dipole traps, however, prevented further exploration within the current work. Nevertheless, we believe our preliminary results reported above bodes well for future extensions to more elaborate geometries and to additional clouds. While the number of BECs handled by our 3D tweezer platform in the present study were limited to four and their purity modest, our previous work \cite{Deb2014} on 1D arrays indicates that advancements to more and pure BECs should be reasonably straightforward using an efficient runaway evaporation scheme for condensing clouds.

In a previous study from our group\cite{Roberts2014}, an optical potential set up by a single, static beam crossed by multiplexed steerable vertical tweezer beams established a linear array of $32$ time averaged optical dipole traps that was filled evenly with utracold atoms (around $10^5$ atoms per well at temperture $\sim 1\mu$K). Ultimately, the number of individual time averaged optical traps that we can achieve with our system is limited by both the available optical power and the maximum rate at which we can toggle frequencies inside an AOD.  In our system the time it takes an acoustic wave to propagate across the input laser beam fixes the maximum frequency toggling rate to be $f_{\rm toggle}^{\rm max}\approx 300$ kHz.  If we have $N_{\rm traps}$ individual traps, the toggling rate seen by the atoms will be $f_{\rm toggle}^{\rm max}/N_{\rm traps}$, and this value must be much larger than twice the harmonic trapping frequency experienced by the atoms to avoid parametrically heating the atoms.  For harmonic trapping frequencies of $\approx 200$ Hz, the above timing considerations imply that we are limited to $N_{\rm traps} \ll 750$ for each pair of AODs if we are not limited in the available optical power.  However, assuming an optical power of at most $\sim$10 W going to each AOD, the above estimate is reduced to $\lesssim$100 individual traps for a 40 $\mu$m beam waist (again assuming a harmonic trapping frequency of 200 Hz for $^{87}$Rb atoms).  Additional timing considerations, such as those discussed in \cite{Bell2018a}, may reduce the maximum achievable number of time averaged traps that we can produce.

While the optical tweezer platform presented in this article allows for elaborate arrangements of dipole traps in three dimensions, it cannot ``paint'' completely arbitrary 3D potentials. The laser beams continue to propagate after forming a crossed dipole trap which, on one hand, opens up the possibility that separate traps can share a beam (see Fig.~\ref{fig1}, where one of the vertical beams is shared by two clouds) implying an efficient use of optical power. On the other hand, this reduces the flexibility for manipulating such traps completely independently and, in general, it means that one must consider the propagation of all other beams when deciding where to place a trap. Moreover, it excludes the possibility of creating certain potentials.

When imaging three-dimensional arrangements of atoms, it is important to remember that the standard technique of absorption imaging gives the atomic density integrated along the imaging axis, i.e. a projection. Therefore, it is necessary to capture at least two images on orthogonal planes to subsequently reconstruct the three-dimensional arrangement, as we are able to do in our configuration. We note that the clouds could be simultaneously imaged along the two axes by employing fluorescence imaging for one axis; in fact, as one of our CCD cameras is of the electron-multiplying type this
mode of operation is accessible to us and will be explored in future work.
\section{Conclusion}
\label{sec:summary}
In this article, we have presented a three-dimensional optical tweezer system based on two pairs of orthogonal AODs. One pair of AODs was used to displace horizontally propagating laser beams in a vertical plane while the other pair was used to displace vertically propagating laser beams in a horizontal plane. Multiple time averaged cross beam dipole traps were simultaneously produced through rapid toggling of the frequency of the sound waves in the AODs. We have demonstrated production of arrays of ultracold atomic clouds in both horizontal and vertical planes. In particular, we have presented a square array of four BECs in a vertical plane and a $2\times2$ array of thermal clouds in a horizontal plane.

In future experiments, the three-dimensional optical tweezer system presented here could be used for creating time averaged optical ring traps~\cite{Bell2016,Bell2018a} and for studying vortex and superfluid flow effects resulting from the merging and collision of independent BECs~\cite{Scherer2007,Mawson2015,Aidelsburger2017}. The dynamic reconfigurability of the optical tweezer system also makes it a candidate as a platform for quantum simulation with engineered Hamiltonians~\cite{Hayward2012,Fialko2017,Sturm2017}.

A recent experiment by Pu~\textit{et al.} has demonstrated the use of multiplexed AODs to demonstrate 22-partite entanglement in a $5\times5$ array of individually accessible atomic quantum interfaces~\cite{Pu2018}. The experiment of Pu~\textit{et al.} used atoms in a single magneto-optical trap. Applying our apparatus to the experimental techniques of Pu~\textit{et al.} would allow entanglement to be demonstrated between truly isolated atomic ensembles.

\begin{acknowledgments}
We thank Wolfgang Wieser for supplying custom firmware for the FlexDDS allowing us to operate it outside of the standard specifications. CSC thanks Mark Baker for useful discussions on magnetic levitation.
ABD was supported by the Marsden Fund of New Zealand (contract UOO1729).

\end{acknowledgments}

%

 \end{document}